\documentclass[letterpaper, preprint, paper,11pt]{AAS}	

\usepackage{bm}
\usepackage{amsmath}
\usepackage{subfigure}
\usepackage{graphicx}
\usepackage[colorlinks=true, pdfstartview=FitV, linkcolor=black, citecolor= black, urlcolor= black]{hyperref}
\usepackage{overcite}
\usepackage{footnpag}			      	

\usepackage{multirow}
\usepackage{multicol}

\usepackage[normalem]{ulem}
\usepackage{cancel}

\usepackage{soul} 

\PaperNumber{23-199}

\begin{document}

\title{Genetic Fuzzy System-based Control for Final Approach of Spacecraft Rendezvous and Proximity Operations}

\author{Daegyun Choi\thanks{PhD Student, Department of Aerospace Engineering and Engineering Mechanics, University of Cincinnati, Cincinnati, OH 45221, USA.}, Donghoon Kim\thanks{Assistant Professor, Department of Aerospace Engineering and Engineering Mechanics, University of Cincinnati, Cincinnati, OH 45221, USA.}, and Henzeh Leeghim\thanks{Professor, Department of Aerospace Engineering, Chosun University, Gwangju 61452, South Korea.}
}

\maketitle{}

\begin{abstract}
In-space servicing has been receiving great attention to extend the operation of spacecraft with defective components. This requires rendezvous and proximity operations for a chaser to provide service to a target. This work constructs a fuzzy inference system-based controller for the chaser to reach the cooperative target on a circular orbit in the final approach phase while minimizing the energy consumption of the chaser. The offline training process performed by a genetic algorithm deals with multiple initial relative positions of the chaser, and the trained controller is validated using a testing environment with disturbances, which differs from the training scenarios.
\end{abstract}

\section{Introduction}
\begin{figure}[!t]
    \centering
    \subfigure[Phasing]{
    \includegraphics[width=0.34\columnwidth]{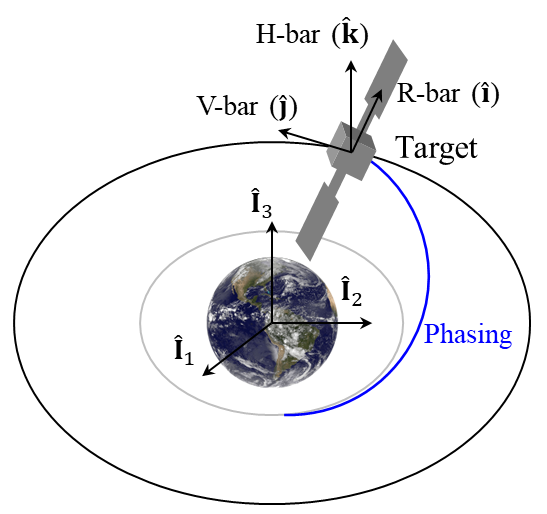}}
    \subfigure[Rendezvous]{
    \includegraphics[width=0.64\columnwidth]{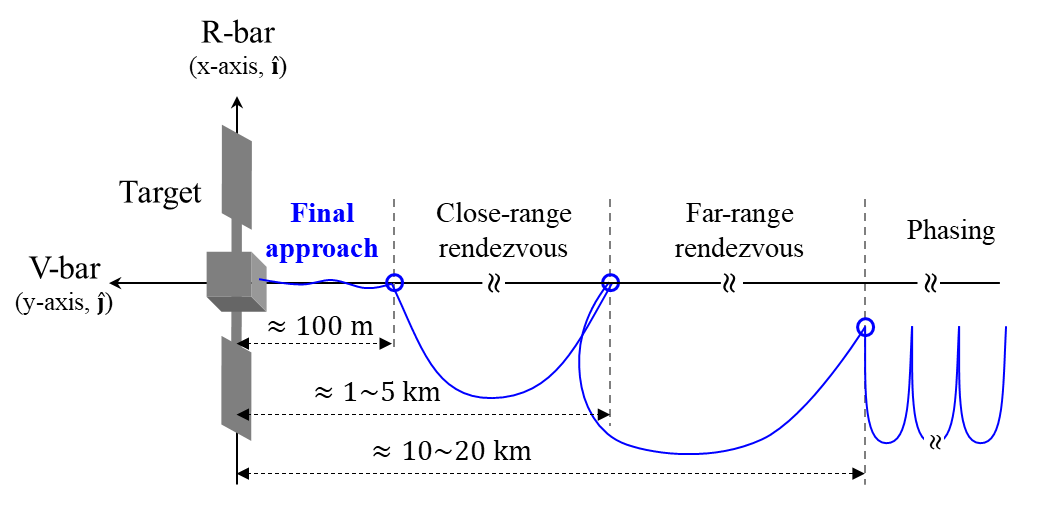}}
    \caption{Phasing and Rendezvous Process}
    \label{fig:rendezvous}
\end{figure}

A great number of spacecraft have been launched into space since the beginning of spaceflight in the 20th century. The expected operation life of the spacecraft usually depends on its mission design. 
However, until now, it has been difficult to carry out any given mission as planned, as it is very hard to repair faulty components in space if a launched spacecraft has one or more malfunctioning sensors and/or actuators.
Recently, in-space servicing, such as refueling, upgrading, and repairing some of the components, has been receiving great attention to extend the mission of spacecraft with faulty components thanks to the advancement of space technologies, such as electronics, rockets, and spacecraft. To perform such missions, rendezvous and proximity operations (RPOs) for servicing spacecraft, named a chaser, are required \cite{Guang2018,Shi2018}. When a chaser approaches the target for servicing from a different orbit, the chaser first performs a phasing process to enter the target's orbit as shown in Figure \ref{fig:rendezvous}. Once the chaser is sufficiently close to the target 
(within about 20 km), the chaser conducts a far- and close-range rendezvous process until the relative distance is less than about 100 m. Then, the chaser finally performs the final approach process to reduce the relative distance between the chaser and the target to {near} zero. Among the entire rendezvous process, this work mainly focuses on the final approach phase.

Many researchers have studied to support safe RPO missions. Some scholars used a widely used collision avoidance method for autonomous robotic platforms, which is the artificial potential field (APF) approach. With the concept of the APF that can provide collision-free trajectories, various control laws were applied to control the spacecraft at the final approach phase of the RPOs. Others proposed control laws combine the APF with non-singular terminal sliding mode control \cite{Li2019JAE}, sub-optimal sliding mode control \cite{Cao2018}, and backstepping control \cite{Li2019}. With the advancement of computing technologies, some researchers have applied artificial intelligence (AI) techniques to diverse control problems and platforms, such as explainable AI-based spacecraft attitude control \cite{Choi2021ascend}, decentralized control of ground robots for a collaborative task \cite{Choi2021electronics}, and collision-free navigation of aerial vehicles \cite{Choi2021robotica}. In particular, some studies applied the reinforcement learning (RL) approach to conduct spacecraft proximity operations \cite{Fedrici2021,Oestreich2021}. These studies validated the performance of the RL for spacecraft RPO missions in terms of energy consumption. However, the RL lacks the explainability of both the decision process and resulting outputs. 

This work aims to design a fuzzy inference system (FIS)-applied controller to produce the control signal for the final approach of the chaser. In fact, the FIS-based controller provides transparent and explainable control inputs unlike the RL approach because the decision-making process is composed of linguistic variables and \emph{If-Then} rules \cite{Mamdani1974,Choi2021robotica}. However, the FIS cannot ensure the optimality of the solution because the design parameters of the FIS are usually determined based on the knowledge of the given system. For this reason, the optimizing capability is combined into the FIS using a genetic algorithm (GA), which is called a genetic fuzzy system (GFS). Since the GA has an aggressive search capability and provides a solution close to a global minimum, the GFS is a good approach for obtaining a FIS controller that can produce a near-optimal control signal. For this reason, this paper proposes a FIS-based controller for the chaser to minimize energy consumption during the final approach phase.

\section{Relative Dynamic Model of Spacecraft}
\begin{figure}[!t]
    \centering
    \includegraphics[width=0.5\columnwidth]{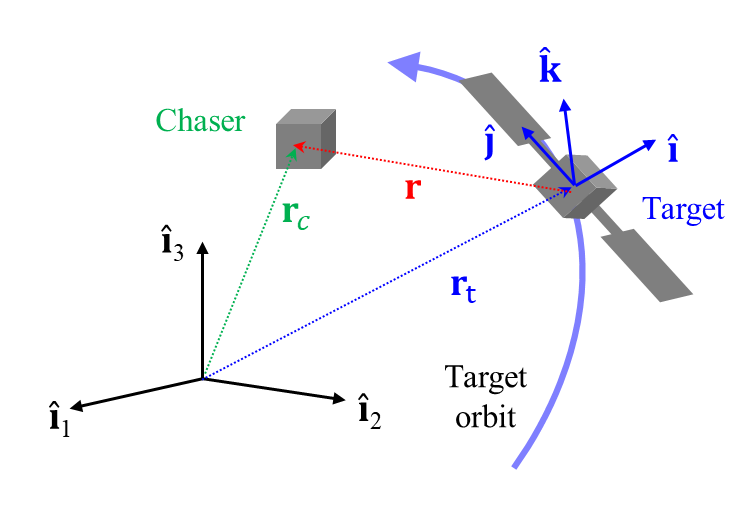}
    \caption{Reference Frames}
    \label{fig:frame}
\end{figure}

This work considers an RPO with two spacecraft orbiting around the Earth, especially for a final approach phase for a successful docking between two spacecraft. Let us consider an active spacecraft, which is a chaser, that actively approaches a passive spacecraft, named a target. One assumes that the target is cooperative, which means the target shares its own state information. Thus, the relative position and velocity of the chaser with respect to the target can be measured via onboard visual sensors. To describe the chaser's translational motion with respect to the target, one introduces a co-moving frame centered on the target. The x-axis (R-bar) is directed from the center of the Earth to the target, and the y-axis (V-bar) is the direction of the target's velocity vector. Therefore, the z-axis (H-bar) is perpendicular to both x- and y-axes, which is the direction of the angular momentum vector of the target's orbit, as shown in Figure \ref{fig:frame}. This co-moving reference frame is used as a local vertical 
local horizontal (LVLH) frame.

With the assumptions that i) the target is on a circular orbit and ii) the relative distance between the target and chase is very small compared to the distance between the center of the Earth and the target, the relative motion of the chaser with respect to the target is described by using the Clohessy-Wiltshire (CW) equation as \cite{Clohessy1960}
\begin{align}
    \ddot{x} -3n^2x -2n\dot{y} & = {(f_x+d_x)}/m ,\label{eq:xddot}\\
    \ddot{y} + 2n\dot{x} & = {(f_y+d_y)}/m ,\label{eq:yddot}\\
    \ddot{z} + n^2z & = {(f_z+d_z)}/m ,
    \label{eq:zddot}
\end{align}
where $x$, $y$, and $z$ are the relative position of the chaser for each axis with respect to the target in the LVLH frame (defined by ${\bf r}=[x,\ y,\ z]^\mathrm{T}$), $f_x$, $f_y$, and $f_z$ are the control force of the chaser for each axis (defined by ${\bf f}=[f_x,\ f_y,\ f_z]^\mathrm{T}$), {$d_x$, $d_y$, and $d_z$ are the disturbance force acting on the chaser (defined by ${\bf d} = [d_x, \ d_y, \ d_z]^\mathrm{T}$),} $m$ is the mass of the chaser, and $n$ is the angular velocity of the target (i.e., mean motion) defined as
$n=\sqrt{{\mu}/{r_\text{t}^2}}$.
Here, $\mu$ is the standard gravitational parameter of the Earth, and $r_\text{t}$ is the distance between the center of the Earth and the target. 
This work assumes that the attitude of the chaser is already synchronized with the attitude of the target to focus on the translational motion of the chaser, and the actuators, such as the thrusters, are aligned to each body axes that can generate the corresponding force. The control input will be determined by the FIS directly, and this will be explained in the following section.

\section{Genetic Fuzzy System-based Control Approach}
\subsection{Genetic Fuzzy System}
\begin{figure}[!t]
    \centering
    \includegraphics[width=0.8\columnwidth]{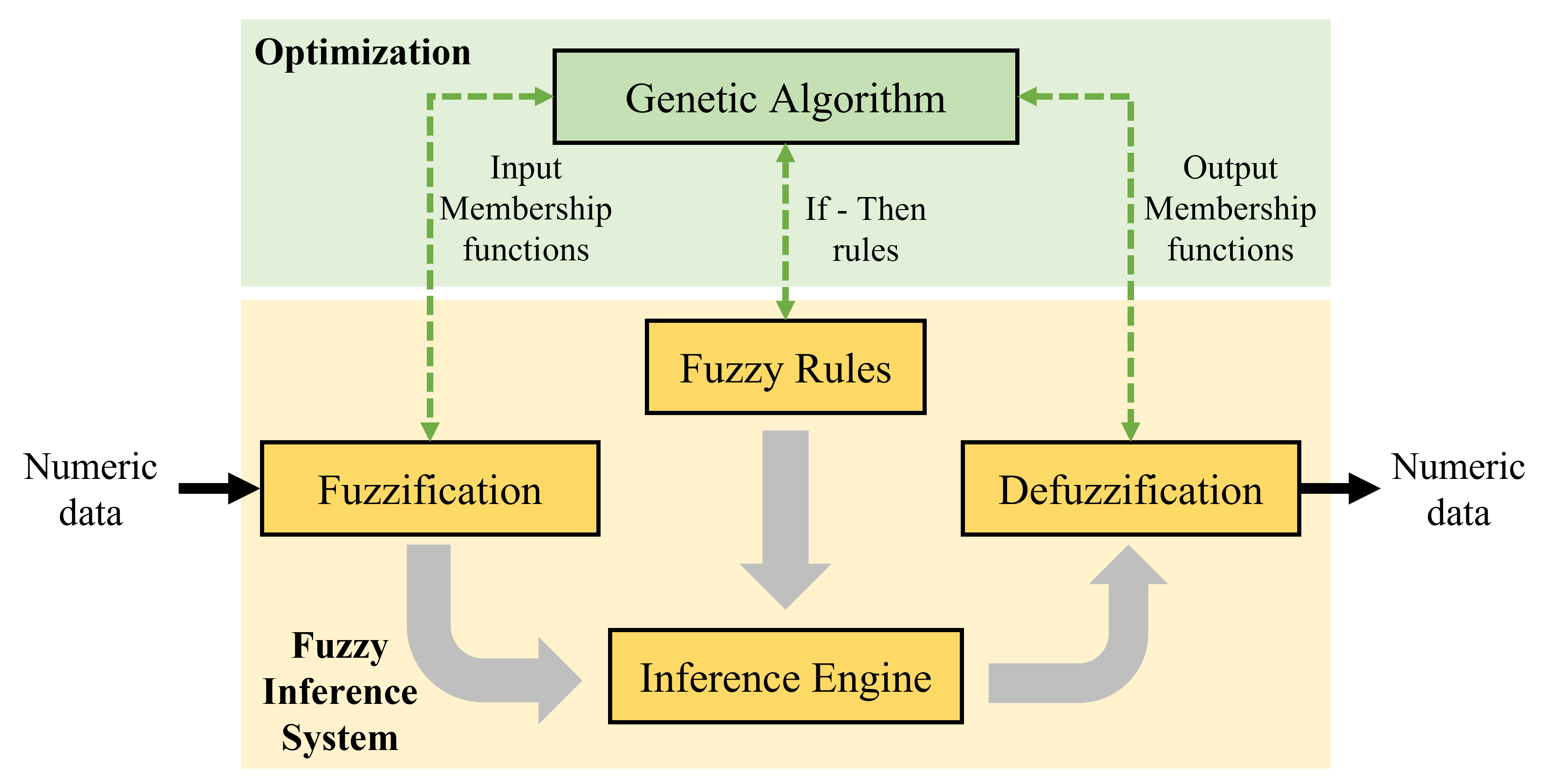}
    \caption{Block Diagram of the Genetic Fuzzy System}
    \label{fig:gfs}
\end{figure}

The FIS is known as a universal approximator \cite{Kosko1994} and utilized as an intelligent control approach \cite{Choi2021electronics} because it has design flexibility and the ability to combine with optimization techniques \cite{Herrera2001}. The FIS is composed of input and output membership functions (MFs) and rules that explain the input and output relationship using \emph{If-Then} statement \cite{Mamdani1974}. Since the decision-making process of the FIS can be described by the rules, the FIS is known as an explainable AI technique. In general, the MFs and the rules are defined by users who have knowledge of the given system. However, this does not guarantee the optimality of the output. Therefore, the FIS can be combined with other approaches to optimize the parameters of the FIS. In particular, this work considers the GA for optimizing the parameters of the MFs and the rules of the FIS \cite{Herrera2001}, which is named as the GFS, as shown in Figure \ref{fig:gfs}. It is important to note that the GA is a metaheuristic optimization technique with certain advantages, such as model-free optimization, fast convergence, and the ability to avoid local minima. By adding the learning capability to the FIS using the GA, the trained FIS can provide a near-optimal output in terms of the defined cost function. After the FIS is trained offline, the trained FIS is applied to generate the control input as a controller.

\subsection{FIS-based Controller Design}
\begin{figure}[!t]
    \centering
    \subfigure[Input 1 - MFs (Relative Position Error)]{\label{fig:mf_input_1}
    \includegraphics[width=0.75\columnwidth]{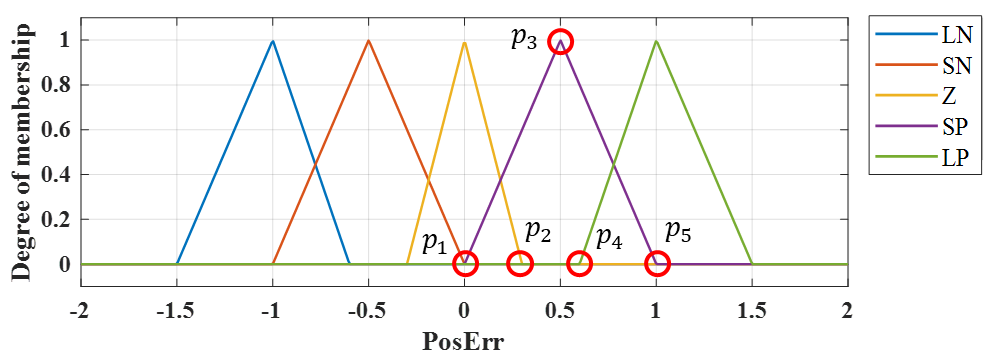}}
    \subfigure[Input 2 - MFs (Relative Velocity Error)]{\label{fig:mf_input_2}   
    \includegraphics[width=0.75\columnwidth]{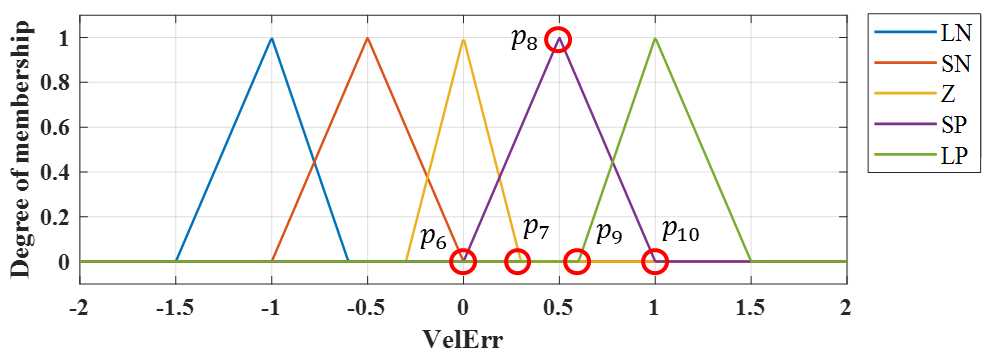}}
    \subfigure[Output MFs (Force)]{\label{fig:mf_output}
    \includegraphics[width=0.75\columnwidth]{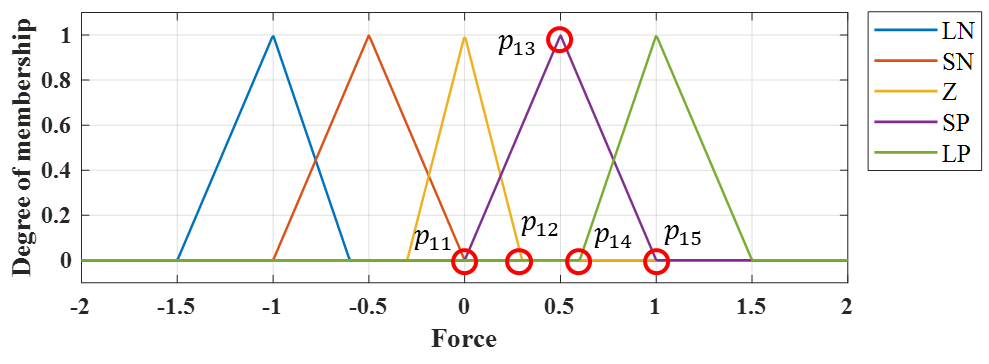}}
    \caption{MFs for the FIS Model}
    \label{fig:mfs}
\end{figure}

This work models a fuzzy controller using the FIS, which has two inputs and one output, to control the translational motion of the chaser for each axis. The FIS is designed to provide the required force which makes the chaser approach the target once the relative position and velocity errors between the chaser and target are given as two inputs. That is, the fuzzy controller is applied to each axis to generate the proper force for the corresponding axis. Usually, depending on the mission, the distance to start the final approach may vary. Therefore, this work utilizes the normalized values for the FIS input and output variables. That is, the minimum and maximum input and output values are set to be -1 and 1, respectively. Especially for the output, the maximum force is multiplied by the output value to convert it into the force. Each MF for input and output has a triangle shape to minimize the complexity as shown in Figure \ref{fig:mfs}. 
Note that each MF represents large negative (LN), small negative (SN), zero (Z), small positive (SP), and large positive (LP), respectively. Here, some of the edges and centers of the input and output MFs are stored into a vector ${\bf p}=[p_1 \ \dots \ p_k]^\mathrm{T}$, where $k$ is the number of parameters to be optimized by the GA. By symmetrically determining the MFs' parameters on the negative side, the number of parameters to be optimized can be reduced.
In a similar manner, some of the rules are also assigned and stored into ${\bf p}$, as shown in Table \ref{tab:rules}. 
\begin{table}[!t]
    \centering
    \caption{Rule-Base for the FIS Model}%
    \label{tab:rules}
    \begin{tabular}{c | c c c c c c}
        \hline
         & & \multicolumn{5}{c}{Relative position error}\\\hline
         & & {LN} & {SN} & {Z} & {SP} & {LP}\\
        \multirow{5}{*}{\shortstack{Relative velocity\\error}} & {LN} & $p_{16}$ &  &  &  & \\
        &{SN} & $p_{17}$ & $p_{21}$ &  \multicolumn{3}{r}{Symmetric} \\
        &{Z} & $p_{18}$ & $p_{22}$ & $p_{25}$ &  &  \\
        &{SP} & $p_{19}$ & $p_{23}$ & $p_{26}$ & $p_{28}$ & \\
        &{LP} & $p_{20}$ & $p_{24}$ & $p_{27}$ & $p_{29}$ & $p_{30}$\\
        \hline
    \end{tabular}
\end{table}
Table \ref{tab:rules} shows the combination of the rules that maps the respective inputs to the corresponding output. The rules can be represented linguistically in terms of the MFs. For instance, let us assume that $p_{25}$ is ``Z''. If the relative position error is ``Z'' and the relative velocity error is ``Z'', the force output is ``Z''. 
Here, the rules are defined in terms of numerical values, each corresponding to an output MF (i.e., 1 refers to LN, 2 refers to SN, 3 refers to Z, and so on.). Note that only the bottom left 15 rules are optimized to reduce the number of parameters to be optimized. 
The stored parameters in ${\bf p}$, which contains 30 components, are then passed onto the GA. In fact, storing all the parameters that need to be optimized in a single vector allows for an easy optimization process.

\subsection{Optimization of Controller}
In the optimization process, the GA optimizes the values in ${\bf p}$. The GA is composed of biologically inspired techniques, such as selection, crossover, mutation, evaluation, and elitism processes. After performing the selection, crossover, and mutation processes, one evaluates the fitness values to find the best-fit solution using the pre-defined fitness function. Next, the elitism process is used to preserve the best solutions at each generation. Note that the most important part of the GA is how the fitness function is determined because it defines the criteria of optimization. In this work, the fitness function is defined as
\begin{equation}
    \mathcal{J} = \int_{0}^{t_{\text{f}}}{\bf f}^\mathrm{T}{\bf f}\ \text{d}t + \rho.
\end{equation}
Here, the defined fitness function is the total control force applied to the chaser over time to reach the target. Also, the fitness function includes a penalty $\rho$ for undesirable behaviors, such as sudden changes in the control force and high velocity near the target. If the undesirable behavior occurs, a large penalty value is applied. By adding the penalty to the fitness function, the best solution avoids undesirable situations while minimizing the fitness function, which is the primary objective.

\section{Simulation Study}
To validate the proposed approach, the numerical simulations are performed using the simulation parameters for the spacecraft and the GA that are listed in Tables \ref{tab:sim_parameters} and \ref{tab:ga_parameters}.
\begin{table}[!t]
    \centering
    \caption{Simulation Parameters for Spacecraft}
    \label{tab:sim_parameters}
    \begin{tabular}{c|c}
        \hline
        Parameters & Values \\ \hline
        Simulation time ($t_\text{f}$) & 200 (s)\\
        Time interval ($\mathrm{d}t$) & 0.1 (s)\\ 
        Chaser's mass ($m$) & 10 (kg) \\
        Target's orbit radius from the center of the Earth ($r_\text{t}$) & 6678 (km)\\
        Maximum force & 0.5 (N) \\
        \hline
    \end{tabular}
\end{table}
\begin{table}[!t]
    \centering
    \caption{Simulation Parameters for the GA}
    \label{tab:ga_parameters}
    \begin{tabular}{c|c}
        \hline
        Parameters & Values \\ \hline
        Maximum number of generations & 300\\
        Population size & 60\\
        Stall number of generations & 100\\
        Selection algorithm & Tournament selection\\
        Tournament size & 4\\
        Crossover algorithm & Two-point crossover\\
        Probability of crossover & 0.85\\
        Mutation algorithm & Adaptive feasible\\
        Probability of mutation & 0.2\\
        Elitism ratio  & 0.05 \\
        \hline
    \end{tabular}
\end{table}

The fuzzy controller proposed is designed to control the chaser at the final approach phase. To develop a robust controller, the parameters of the fuzzy controller are optimized in the training process by the GA with multiple training scenarios having different initial relative positions of the chaser from the target ($\pm5$, $\pm50$, and $\pm100$ m) without considering disturbances. In fact, to ensure coverage of various relative positions ranges from negative to positive and from small to large values, combinations of the initial relative positions are used in the training process. In addition, to avoid undesirable behaviors of the chaser, {a} penalty value {greater than the total control force, for example, 100 in this work,} is applied when the conditions listed in Table \ref{tab:penalty} are not met. In particular, the first penalty condition is employed to smoothly approach the target when the chaser is close to the target. The second condition is related to the generation of the smooth control force.
\begin{table}[htbp]
    \centering
    \caption{Required Conditions Avoiding Penalty}
    \label{tab:penalty}
    \begin{tabular}{c|c}
    \hline
        Penalty parameters & Conditions\\ \hline
        Average of the norm of the relative velocity error during the last 30 s & $\le$ 0.05 m/s\\
        Maximum absolute value of the time derivative of the control force& $\le$ 0.1 N/s \\
        \hline
    \end{tabular}
\end{table}

\begin{figure}[!ht]
    \centering
    \subfigure[Relative Position]{\label{fig:train_pos}
    \includegraphics[width=0.85\columnwidth]{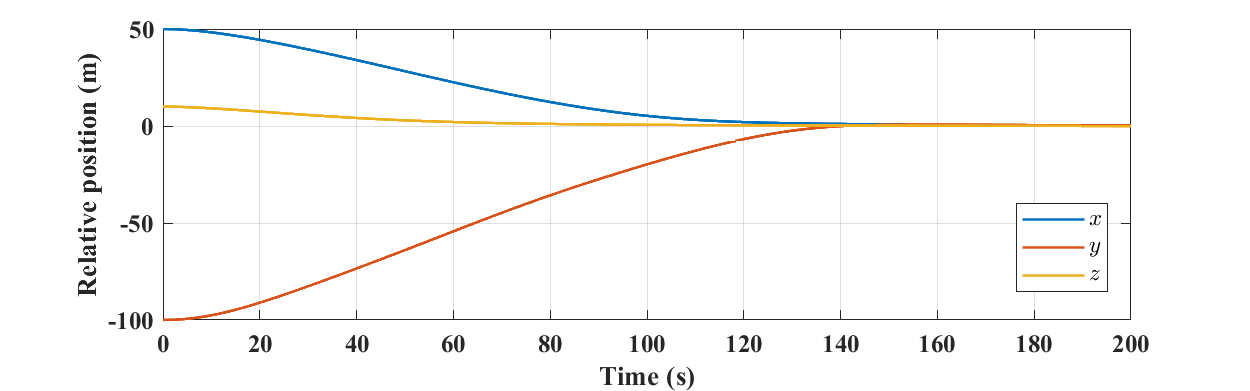}}
    \subfigure[Relative Velocity]{\label{fig:train_vel}   
    \includegraphics[width=0.85\columnwidth]{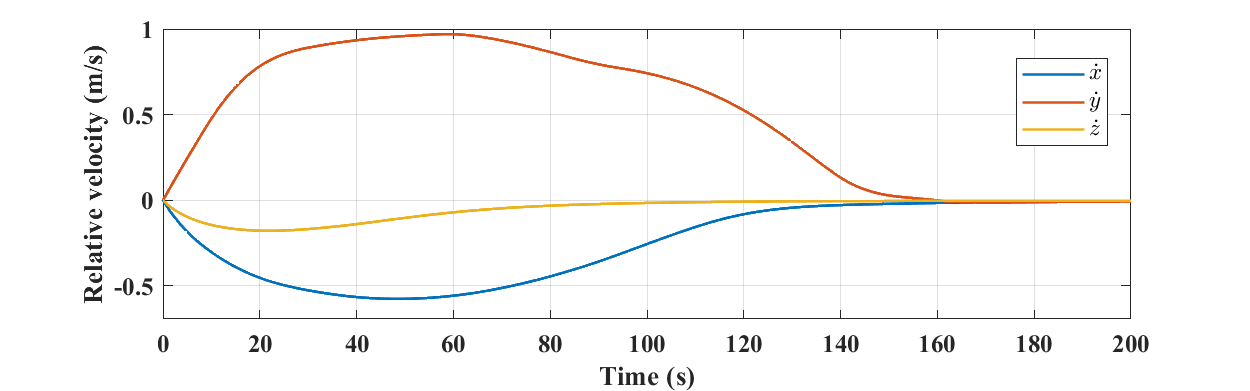}}
    \subfigure[Force]{\label{fig:train_force}
    \includegraphics[width=0.85\columnwidth]{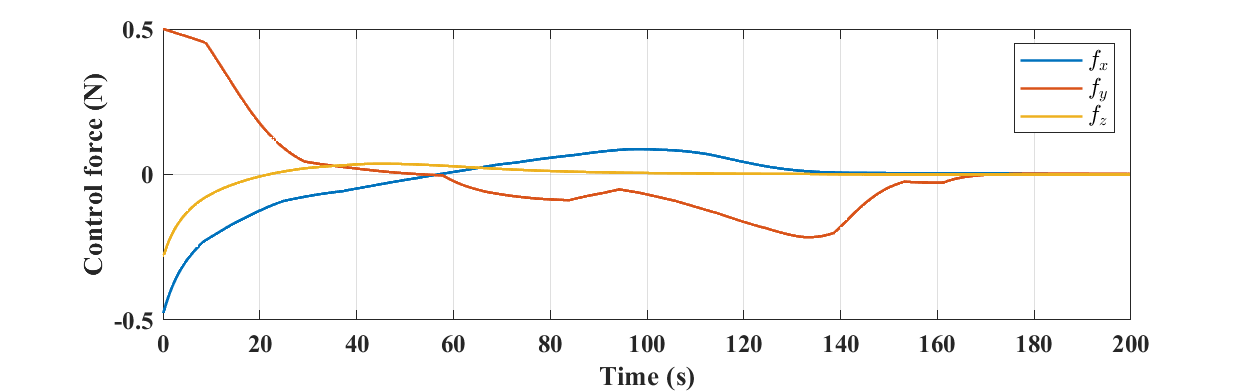}}
    \caption{Chaser's State (Training)}
    \label{fig:train}
\end{figure}

One can obtain the trained FIS that can generate near-minimal control force after the training process. Among several training scenarios, Figure \ref{fig:train} shows one of the results, which contains the time history of the relative position, relative velocity, and force of the chaser. It is shown that all the states are converted to zero, and one can realize that the chaser smoothly approaches the target during maneuver even though the target is also orbiting around the Earth. Since the maximum force is limited to 0.5 N, the control force shown in Figure \ref{fig:train_force} does not exceed the maximum force and smoothly converges to zero. In fact, the energy optimal solution, which satisfies the given initial and final states, is usually obtained as bang-off-bang control, and the optimal control at the final time is not zero. Since the chaser is very close to the target right after the final time, non-zero control force may cause dangerous situations. Thus, this work considers producing smooth control trajectory to avoid such risk by adding the second penalty condition listed in Table \ref{tab:penalty}. It is obtained that the average of the norm of the relative velocity error during the last 50 seconds is 0.015 m/s and the maximum value of the time derivative of the control force is 0.0261 N/s, which satisfy the conditions avoiding the penalty. 

Although the training process is performed without considering disturbances, the testing scenario considers external disturbances around the chaser in order to validate the performance of the trained fuzzy controller in the presence of uncertainties. To confirm the convergence trend of the chaser's position trajectories, the simulation time is increased in the testing scenario. Also, the initial relative position and the disturbance considered including the simulation time are listed in Table \ref{tab:test_parameters}.

\begin{figure}[!ht]
    \centering
    \subfigure[Relative Position]{\label{fig:test_pos}
    \includegraphics[width=0.85\columnwidth]{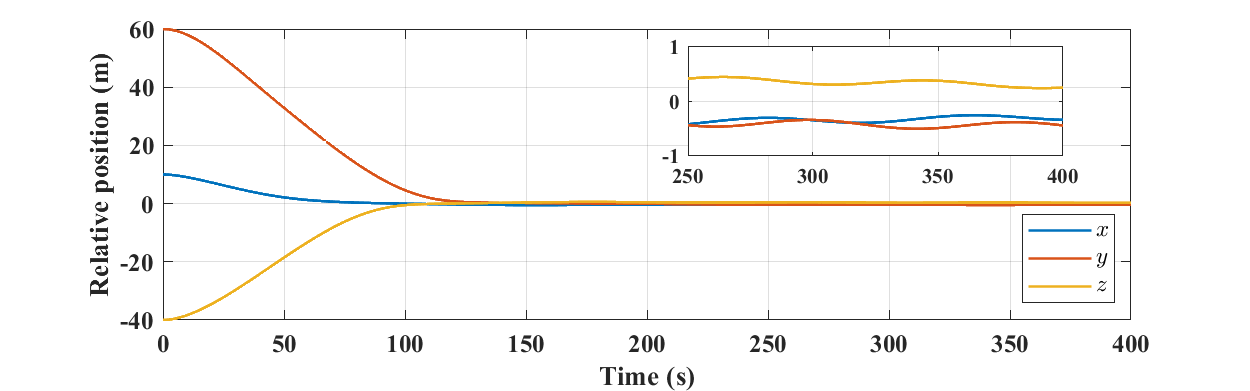}}
    \subfigure[Relative Velocity]{\label{fig:test_vel}   
    \includegraphics[width=0.85\columnwidth]{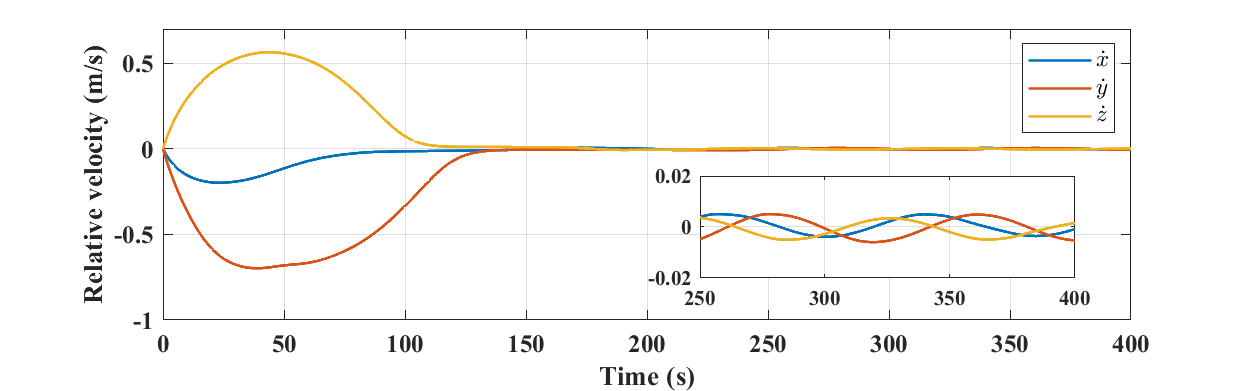}}
    \subfigure[Force]{\label{fig:test_force}
    \includegraphics[width=0.85\columnwidth]{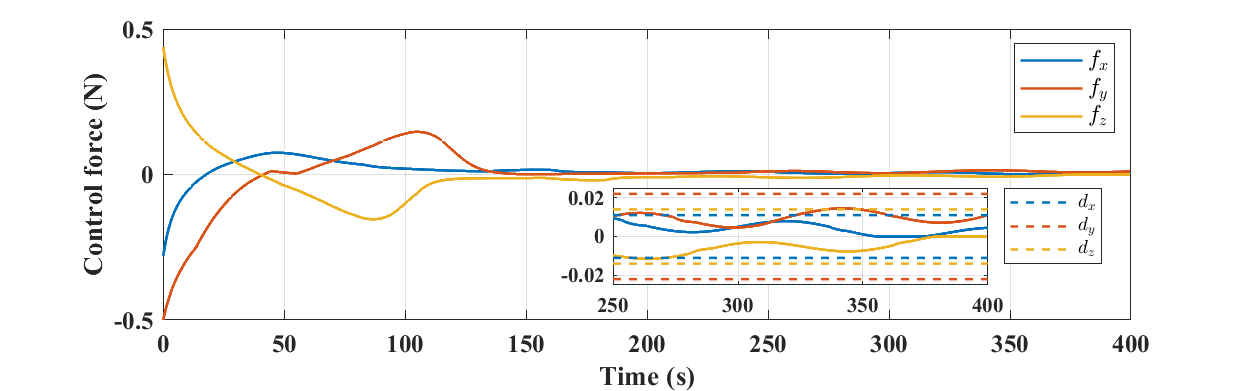}}
    \caption{Chaser's State (Testing)}
    \label{fig:test}
\end{figure}

The testing results shown in Figure \ref{fig:test} display that the trained fuzzy controller provides the proper control input that makes the chaser approach the target even though the disturbance force is applied to the chaser. The relative position of the chaser converges within $\pm$ 0.5 m during the last 200 seconds, and the relative velocity of the chaser is also reduced, which is less than $\pm$ 0.01 m/s. In addition, the generated control force remains very small, which is less than the level of the disturbances. This result shows that the chaser can reach the target smoothly in the presence of the disturbance force using the trained fuzzy controller proposed. It can be seen that the FIS-based controller is robust to uncertainties even though the disturbances are not included in the training process. 

\begin{table}[!t]
    \centering
    \caption{Simulation Parameters for Testing}
    \label{tab:test_parameters}
    \begin{tabular}{c|c}
    \hline
        Parameters & Values \\ \hline
        Simulation time & 400 (s) \\
        Initial relative position &  $\begin{bmatrix} 10 & 60& -40\end{bmatrix}^\mathrm{T}$ (m) \\
        Disturbance (${\bf d}$) & $
        \begin{bmatrix}-1.1\cos \pi nt & -2.2 \sin\pi nt+\pi /3 & 1.4 \cos\pi nt\end{bmatrix}^\mathrm{T}
        \times 10^{-2}$ (N) \\ \hline
    \end{tabular}
\end{table}

\section{Conclusions}
This work proposes a fuzzy inference system-based control model for a chaser to reach a cooperate target on a circular orbit in the final approach phase while minimizing the energy consumption of the chaser. The proposed fuzzy controller provides the proper force required using the relative position and velocity error. Also, the proposed fuzzy controller is trained by a genetic algorithm. The offline training process deals with multiple initial relative positions of the chaser while the trained controller is applied to a testing environment including disturbances, which is different from the training scenarios. Consequently, the results showed that the proposed fuzzy controller properly produces the control force for the chaser to smoothly approach and reach the target. In future work, the proposed controller will be improved by {conducting more training to obtain better control force trajectories, and} a collision avoidance capability that can safely navigate around the target {will be added to the controller}.




\bibliographystyle{AAS_publication}   
\bibliography{references}   

@article{Guang2018,
    author = {Guang, Zhai and Heming, Zheng and Liang, Bin},
    title = {Attitude Dynamics of Spacecraft with Time-Varying Inertia During On-Orbit Refueling},
    journal = {Journal of Guidance, Control, and Dynamics},
    volume = {41},
    number = {8},
    pages = {1744-1754},
    year = {2018},
    doi={10.2514/1.G003474}
}

@ARTICLE{Shi2018,
    author={Shi, Lingling and Katupitiya, Jayantha and Kinkaid, Nathan Michael},
    journal={IEEE Transactions on Aerospace and Electronic Systems},
    title={Hybrid Control of Space Robot in On-Orbit Screw-Driving Operation},
    year={2018},
    volume={54},
    number={3},
    pages={125doi=3-1264},
    doi={10.1109/TAES.2017.2780558}
}

@article{Li2019,
    title = {Artificial potential field based robust adaptive control for spacecraft rendezvous and docking under motion constraint},
    journal = {ISA Transactions},
    volume = {95},
    pages = {173-184},
    year = {2019},
    issn = {0019-0578},
    doi = {https://doi.org/10.1016/j.isatra.2019.05.018},
    author = {Qi Li and Jianping Yuan and Bo Zhang and Huan Wang}
}

@article{Li2019JAE,
    author = {Xuehui Li and Zhibin Zhu and Shenmin Song},
    title ={Non-cooperative autonomous rendezvous and docking using artificial potentials and sliding mode control},
    journal = {Proceedings of the Institution of Mechanical Engineers, Part G: Journal of Aerospace Engineering},
    volume = {233},
    number = {4},
    pages = {1171-1184},
    year = {2019},
    doi = {10.1177/0954410017748988}
}

@article{Cao2018,
    title = {Suboptimal artificial potential function sliding mode control for spacecraft rendezvous with obstacle avoidance},
    journal = {Acta Astronautica},
    volume = {143},
    pages = {133-146},
    year = {2018},
    issn = {0094-5765},
    doi = {https://doi.org/10.1016/j.actaastro.2017.11.022},
    author = {Lu Cao and Dong Qiao and Jingwen Xu}
}

@article{Choi2021robotica,
    title={Intelligent cooperative collision avoidance via fuzzy potential fields},
    DOI={10.1017/S0263574721001454},
    journal={Robotica},
    publisher={Cambridge University Press},
    author={Choi, Daegyun and Chhabra, Anirudh and Kim, Donghoon},
    year={2021},
    pages={1–20}
}

@Article{Choi2021electronics,
    AUTHOR = {Choi, Daegyun and Kim, Donghoon},
    TITLE = {Intelligent Multi-Robot System for Collaborative Object Transportation Tasks in Rough Terrains},
    JOURNAL = {Electronics},
    VOLUME = {10},
    YEAR = {2021},
    NUMBER = {12},
    ARTICLE-NUMBER = {1499},
    DOI = {10.3390/electronics10121499}
}

@inproceedings{Choi2021ascend,
    author = {Daegyun Choi and Anirudh Chhabra and Donghoon Kim},
    title = {Genetic Algorithm-Aided Fuzzy Controller for Spacecraft Attitude Maneuver with Uncertainties},
    booktitle = {ASCEND 2021},
    chapter = {},
    pages = {},
    year = {2021},
    doi = {10.2514/6.2021-4174}
}

@article{Fedrici2021,
    author = {Federici, Lorenzo and Benedikter, Boris and Zavoli, Alessandro},
    title = {Deep Learning Techniques for Autonomous Spacecraft Guidance During Proximity Operations},
    journal = {Journal of Spacecraft and Rockets},
    volume = {58},
    number = {6},
    pages = {1774-1785},
    year = {2021},
    doi = {10.2514/1.A35076}
}

@article{Oestreich2021,
    author = {Oestreich, Charles E. and Linares, Richard and Gondhalekar, Ravi},
    title = {Autonomous Six-Degree-of-Freedom Spacecraft Docking with Rotating Targets via Reinforcement Learning},
    journal = {Journal of Aerospace Information Systems},
    volume = {18},
    number = {7},
    pages = {417-428},
    year = {2021},
    doi = {10.2514/1.I010914}
}

@book{Herrera2001,
    author = {Herrera, F. and Hoffmann, F. and Magdalena, L.},
    title = {Genetic Fuzzy Systems: Evolutionary Tuning and Learning of Fuzzy Knowledge Bases},
    publisher = {Singapore: World Scientific Publishing Company},
    year = {2001}
}

@article{Clohessy1960,
    author = {CLOHESSY, W. H. and WILTSHIRE, R. S.},
    title = {Terminal Guidance System for Satellite Rendezvous},
    journal = {Journal of the Aerospace Sciences},
    volume = {27},
    number = {9},
    pages = {653-658},
    year = {1960},
    doi = {10.2514/8.8704}
}

@article{Mamdani1974,
  title={Applications of fuzzy algorithms for control of a simple dynamic plant},
  author={Ebrahim H. Mamdani},
  journal={Proceedings of the IEEE},
  year={1974}
}

@article{Kosko1994,
    author={Kosko, B.},
    journal={IEEE Transactions on Computers},   title={Fuzzy systems as universal approximators},
    year={1994},
    volume={43},
    number={11},
    pages={1329-1333},
    doi={10.1109/12.324566}
}

\end{document}